\newcommand{\ba}{\begin{array}}
\newcommand{\ea}{\end{array}}
\newcommand{\bea}{\begin{eqnarray}}
\newcommand{\eea}{\end{eqnarray}}
\newcommand{\be}{\begin{equation}}
\newcommand{\ee}{\end{equation}}
\newcommand{\gapproxeq}{\lower .7ex\hbox{$\;\stackrel{\textstyle >}{\sim}\;$}}
\newcommand{\lapproxeq}{\lower .7ex\hbox{$\;\stackrel{\textstyle <}{\sim}\;$}}

\documentstyle[11pt]{article}
\def\ep{\epsilon}
\def\epp{\epsilon'}
\def\ept{\tilde\epsilon}
\def\ks{K_S}
\def\kl{K_L}
\def\ko{K^0}
\def\kbo{\overline{K}^0}
\def\ku{K_1}
\def\kd{K_2}
\def\f{\pi^+\pi^-\pi^0}
\def\riar{\rightarrow}

\begin{document}
\baselineskip=18pt

{\hfill \bf DSF preprint 93/52~~~~~~~~}

{\hfill \bf SUHEP preprint 582 ~~~~~~~~}

\vskip 1.0truecm
\centerline{\bf Energy-Charge Correlation in the $\f$ Decay of $\kl$ and of
Tagged Neutral Kaons}

\vskip 1cm
\centerline{ F. Buccella$^a$, O. Pisanti$^{b,c}$
\footnote{{\it e-mail:}~pisanti@na.infn.it}, F. Sannino$^{b,c,d}$
\footnote{{\it e-mail:}~sannino@na.infn.it, sannino@suhep.phy.syr.edu} }
\vskip 1cm

\begin{itemize}
\item[$^a$] {\it Istituto di Fisica Teorica, Mostra d'Oltremare, 80125 Napoli,
Italia.}
\item[$^b$] {\it Dipartimento di Scienze Fisiche, Mostra d'Oltremare Pad.19,
80125 Napoli, Italia.}
\item[$^c$] {\it Istituto Nazionale di Fisica Nucleare, Sezione di Napoli,
Mostra d'Oltremare, Pad.19 I-80125 Napoli, Italia.}
\item[$^d$] {\it Department of Physics, Syracuse University, Syracuse, New
York, 13244-1130.}
\end{itemize}

\vskip 3cm
\begin{abstract}
We relate the asymmetries in the charged pions energy in the decay into
$\f$ of $\kl$ and of the tagged neutral kaons. The former asymmetry is a
given combination of $\Re (\ep)$, $\Im (\ep)$, and $|\epp|$. Moreover, the
non-violating CP asymmetry allows a test for the $\chi$PT predictions
within the Zel'dovich approach for the final state interaction.
\end{abstract}

\newpage

\section{Introduction}

\vskip 0.5cm

\indent

The kaon system has been the most natural laboratory to study the CP violation
since its first evidence in $\kl\riar \pi\pi$ decays \cite{chris}.

The phenomenology about CP violation in the $K\riar \pi\pi$
can be described by the well known parameters $\ep$ and $\epp$ \cite{wolf},
\bea
\ep &=& \ept+i \frac{\Im (A_0)}{\Re (A_0)}, \\
\epp &=& \frac{i}{\sqrt{2}}
 e^{i(\delta_2-\delta_0)} \frac{\Re (A_2)}{\Re (A_0)}
\left( \frac{\Im (A_2)}{\Re (A_2)} - \frac{\Im (A_0)}{\Re (A_0)} \right), \\
i~A_I &\equiv& <(\pi \pi)_I | H_W | \ko >,
\eea
where $\delta_2$ and $\delta_0$ are the $\pi \pi$ phase shifts \cite{pdg},
\be
\delta_2-\delta_0+\frac{\pi}{2} = (47\pm 5)^\circ. \label{eq:fasi}
\ee

The parameter $\ep$ is connected with the CP violation in the mass matrix
of the kaons through $\ept$, which appears in the expressions of the
mass eigenstates $\kl$ and $\ks$,
\bea
\ks &=& \frac{(1+\ept)\ko+(1-\ept)\kbo}{\sqrt{2(1+|\ept|^2)}}
=\frac{\ku+\ept \kd}{\sqrt{1+|\ept|^2}}, \label{eq:kappas} \\
\kl &=& \frac{(1+\ept)\ko-(1-\ept)\kbo}{\sqrt{2(1+|\ept|^2)}}
=\frac{\kd+\ept \ku}{\sqrt{1+|\ept|^2}}, \label{eq:kappal}
\eea
where $\ku$ and $\kd$ are respectively the CP even and odd eigenstates, and
$\kbo= CP\ko$.
On the other side, $\epp$ is due to the  CP violating term in the kaon decay
matrix.

The two CP violating parameters $\ep$ and $\epp$ enter in the following
ratios \cite{wolf}:
\bea
\eta_{+-} &\equiv& \frac{A(\kl \riar \pi^+ \pi^-)}
{A(\ks \riar \pi^+ \pi^-)} \simeq \ep + \epp,
\\
\eta_{00} &\equiv& \frac{A(\kl \riar \pi^0 \pi^0)}
{A(\ks \riar \pi^0 \pi^0)}
\simeq \ep - 2 \epp .
\eea
Due to the phenomenological suppression of the $\Delta I=3/2$, the $\epp$
value is very small ($\lapproxeq 10^{-5}$) and therefore difficult to be
measured experimentally.

The "golden-ratio"
\be
\frac{\Gamma
[\ks \riar \pi^0\pi^0]  \Gamma [\kl \riar \pi^+\pi^-]}{\Gamma
[\ks \riar \pi^+\pi^-]  \Gamma [\kl \riar \pi^0\pi^0]}
\simeq 1 + 6 \Re \left( \frac{ \epp}{\ep} \right)
\ee
has been used in measuring $\Re (\epp / \ep)$ by the Collaborations {\it NA31}
at CERN \cite{na31} and {\it E731} at Fermilab \cite{e731}, which gave the
following different results:
\be
\ba{l}
(2.0 \pm 0.7)\cdot 10^{-3}~~~~~~~{\it NA~31} \nonumber \\
(0.74\pm0.60)\cdot 10^{-3}~~~~{\it E~731}. \nonumber
\ea
\label{eq:na31e731}
\ee
These measures require more investigations to better understand the real
underlying mechanism of the direct CP violation.

The two dedicated experiments {\it LEAR} \cite{lear} and {\it DA$\Phi$NE}
\cite{dafne} were planned to improve these measurements and, at the same
time, to enlarge the phenomenology of CP violation in the kaon processes.

The phase of $\ep$, $\phi(\ep)$, is predicted by unitarity to be \cite{pdg}
\be
\phi(\ep) \equiv tan^{-1} \left(\frac{\Im (\ep)}{\Re (\ep)}\right)
= tan^{-1} \left(\frac{2(M_L-M_S)}{\Gamma_S-\Gamma_L}\right)
=(43.68\pm 0.14)^\circ,
\ee
very near to the phase of $\epp$ given by eq.~(\ref{eq:fasi}). So the
result given in eq.~(\ref{eq:na31e731}) indicate that $|\epp|$ is three
order of magnitude smaller than $|\ep|$ and therefore the phase of $\ep$
is, with a good approximation, equal to the phases of $\eta_{+-}$ and
$\eta_{00}$ measured from the interference of $\ks$ and $\kl$ in the decays
into two pions. Indeed, in these experiments one finds \cite{pdg}
\bea
\phi_{+-} &\equiv& tan^{-1} \left(\frac{\Im (\eta_{+-})}{\Re (\eta_{+-})}
\right) = (46.6\pm 1.2)^\circ, \\
\phi_{00} &\equiv& tan^{-1} \left(\frac{\Im (\eta_{00})}{\Re (\eta_{00})}
\right) = (46.6\pm 2.0)^\circ,
\eea
in fair agreement one each other and larger but consistent with the value
predicted by unitarity.

$\Re (\ep)$ may be related to the asymmetry in the semileptonic decays of
$\kl$, where
\bea
\frac{\Gamma (\kl \riar \mu^{+} \pi^{-} \nu_\mu)
-\Gamma (\kl \riar \mu^{-} \pi^{+} \overline{\nu}_\mu)}
{\Gamma (\kl \riar \mu^{+} \pi^{-} \nu_\mu)
+\Gamma (\kl \riar \mu^{-} \pi^{+} \overline{\nu}_\mu)} &=& (0.304\pm
0.025) \%, \label{eq:ramu} \\
\frac{\Gamma (\kl \riar e^{+} \pi^{-} \nu_e)
-\Gamma (\kl \riar e^{-} \pi^{+} \overline{\nu}_e)}
{\Gamma (\kl \riar e^{+} \pi^{-} \nu_e)
+\Gamma (\kl \riar e^{-} \pi^{+} \overline{\nu}_e)} &=& (0.333\pm
0.014) \%. \label{eq:rael}
\eea
The two ratios of eqs.~(\ref{eq:ramu}) and (\ref{eq:rael}), assuming
$\Delta Q=\Delta S$, are both given by $2 \Re (\ep)$, in quite good
agreement with
\be
\Re (\ep) =(1.637\pm 0.013)\cdot 10^{-3},
\ee
found from $|\ep|$ and $\phi(\ep)$.

In this paper we give a model independent way to measure a combination of
$\Re (\ept)=\Re (\ep)$  and $\Im(\ept)=\Im(\ep)-\Im(A_0)/\Re(A_0)$
by relating a particular time-dependent integrated
CP conserving asymmetry, connecting the strangeness of the neutral kaons to
the energy in the Dalitz plot of the charged pions in the decay $\ko (\kbo)
\riar\f$, to the corresponding CP violating term in $\kl\riar\f$.

The time-dependent CP conserving and CP violating asymmetries may be
measured in {\it LEAR} \cite{lear} and {\it DA$\Phi$NE} \cite{dafne}
respectively. The measure of the CP conserving asymmetry will provide a
test of $\chi PT$ (Chiral Perturbation Theory) and give information on the
final state interactions of the three pions.

\section{Asymmetries}

\vskip 0.5cm

\indent

The amplitude for the decay of $\kl\riar\f$ may be given in terms of the
corresponding amplitudes into the same channel of the CP eigenstates $\ku$
and $\kd$, which are,
up to quadratic order in pion energies and disregarding terms with isospin
greater than 3/2 in the weak Hamiltonian \cite{zedema,zeld},
\bea
A(\ku \riar\f) &=& \tilde{a} e^{i \delta_S(1)}
+ \tilde{d} \left(\rho^2 + \frac{\Delta^2}{3}\right)
+ \tilde{b} \rho e^{i \delta_{MS}(1)} + \tilde{e} \left(\rho^2
- \frac{\Delta^2}{3} \right), \label{eq:amk1} \\
& & +i\left[ c \Delta e^{i \delta_{MA}(2)} + f \rho \Delta \right],
 \nonumber \\
A(\kd \riar\f) &=& i\left[ a e^{i \delta_S(1)} + d \left(\rho^2 +
\frac{\Delta^2}{3} \right) \right]
+ i\left[ b\rho e^{i \delta_{MS}(1)} + e \left(\rho^2 - \frac{\Delta^2}{3}
\right) \right] \label{eq:amk2} \\
& & + \tilde{c}\Delta e^{i \delta_{MA}(2)}
+ \tilde{f} \rho \Delta , \nonumber
\eea
where
\be
\Delta \equiv \frac{E_+ - E_-}{m_K},~~~~~~~~~~~\rho \equiv \frac{E_0}
{m_K} - \frac{1}{3}
\ee
are auxiliary variables over the Dalitz plot ($E_i$ is the energy of
$\pi_i$).
In the following we will assume SU(2) flavour symmetry, i.e.
$m_{\pi^\pm}=m_{\pi^0}$.
All the coefficients in eqs.~(\ref{eq:amk1}) and (\ref{eq:amk2}) are
real in our convention.
The CP violating terms are represented by the letters with a tilde;
$\delta_S(1)$ and $\delta_{MS}(1)$ represent the totally symmetric and
mixed symmetric isospin 1 final state interaction, while $\delta_{MA}(2)$ is
connected to the mixed antisymmetric isospin 2 interaction.
The factorization for the amplitudes, in terms of
the pion phase shifts, relies on the Watson's theorem \cite{dona}.
The coefficients of the isospin 2 term are due only to the $\Delta
I=3/2$ operator in the weak interaction, while the other ones are a
combination of $\Delta I=1/2$ and $\Delta I=3/2$.

In the
Standard Model, the direct CP violation in the amplitudes comes from
the diagrams corresponding to the "penguin" operators which transform as
the $ \Delta I = 1/2$, $|\Delta S|=1$, $|\Delta Q|=0$ member
of an octet; this suggests to adopt the $\Im (A_2)=0$ phase convention as
the most natural one.
Moreover, it is justified to eliminate the direct phenomenological
$\Delta I =3/2~~$ CP violating terms, $\tilde{c}$ and $\tilde{f}$,
in the amplitude.
By keeping only the CP conserving part of $A(\ku\riar\f)$ one obtains
\bea
A(\kl \riar\f) &=& i\left[ a e^{i \delta_S(1)} + d \left(\rho^2 +
\frac{\Delta^2}{3} \right) \right]
+ i\left[ b\rho e^{i \delta_{MS}(1)} + e \left(\rho^2 - \frac{\Delta^2}{3}
\right) \right] + \label{eq:amkl} \\
&+& i \ept\left[ c \Delta e^{i \delta_{MA}(2)}+ f \rho \Delta \right],
\nonumber
\eea
{}from which we can derive the CP violating asymmetry
\be
\Gamma_\Delta (\kl) \equiv \frac{F\left[\Delta \cdot
|A(\kl\riar\f) |^2 \right]~F[1]}{\Gamma [\kl \riar \f]~F[|\Delta|]}.
\label{eq:askl}
\ee
$F[...]$ represents the phase space integration defined in the Appendix
B, so that
\be
F[|A(\kl\riar\f)|^2]=\Gamma(\kl\riar\f).
\ee

After some algebra we find
\be
F\left[\Delta \cdot
|A(\kl\riar\f) |^2 \right]=2~[P~\Re(\ept)+S~\Im(\ept)],
\ee
where we define
\be
F[\Delta~A(\ku \riar \f)^{*} A(\kd \riar \f)] \equiv P+i~S.
\label{eq:pes}
\ee
The complete expression for $P$ and $S$ can be found in the Appendix A.
Now, eq.~(\ref{eq:askl}) becomes
\be
\Gamma_\Delta (\kl) = \frac{2~[P~\Re(\ept)+S~\Im(\ept)]
{}~F[1]}{\Gamma [\kl \riar \f]~F[|\Delta|]}. \label{eq:nostra}
\ee

An information on the coefficients of $\Re(\ept)$ and $\Im(\ept)$ in
eq.~(\ref{eq:nostra}) can be obtained by considering the
strangeness-charged pions energy correlation, which may be measured at
{\it LEAR} \cite{lear}, where it is possible to tag the initial strangeness
of the neutral kaons. The latter evolve according to:
\bea
\ko(t) &=& \frac{\sqrt{1+|\ept|^2}}{\sqrt{2}(1+\ept)}
[\ks e^{-{\Gamma_S t \over 2}-iM_S t}+
\kl e^{-{\Gamma_L t \over 2}-iM_L t}], \label{eq:timeko}
\\  & & \nonumber \\
\kbo (t) &=& \frac{\sqrt{1+|\ept|^2}}{\sqrt{2}(1-\ept)}
[\ks e^{-{\Gamma_S t \over 2}-iM_S t}-\kl
e^{-{\Gamma_L t \over 2}-iM_L t}]. \label{eq:timekbo}
\eea
In the previous expressions the standard notation \cite{wolf} has been
adopted.

In order to reach our goal let us consider the CP conserving asymmetry
\cite{sei}
\be
\Sigma^{+-0}_\Delta (T) \equiv \int^T_0
[\Gamma_\Delta (t) - \overline{\Gamma}_\Delta (t)] dt,
\label{eq:sigma}
\ee
where
\bea
\Gamma_\Delta (t) &\equiv& F\left[\Delta \cdot
|A(\ko (t) \riar\f)|^2 \right], \label{eq:gamt}
\\
\overline{\Gamma}_\Delta (t) &\equiv& F\left[ \Delta \cdot
| A(\kbo(t) \riar\f)|^2 \right], \label{eq:gambt}
\eea
and $\int^{T}_{0}\Gamma_{\Delta}(t) (\overline{\Gamma}_{\Delta}(t))~dt$
are reported in the Appendix A.

A straightforward calculation gives the following result \cite{sei}:
\be
\Sigma^{+-0}_\Delta (T) =
f(\overrightarrow{F[\Delta~A(\ku \riar \f)^{*} A(\kd \riar \f)]},T)=
P~f(\vec{1},T)+S~f(\vec{i},T),
\ee
where the $f(\vec{\mu},T)$, reported in two special cases in fig.~1,
are defined in the Appendix A.
So, by studying the time dependence of $\Sigma^{+-0}_\Delta (T)$ we can
extract the coefficients which appear in the expression (\ref{eq:nostra}).
To get an idea of the size of the effect, in the next section we
shall give some theoretical evaluation within the framework of
Chiral Perturbation Theory ($\chi$PT).

\begin{figure}
\vspace{5cm}
\caption{The behaviour of $f(\vec{1},T)$ (full line) and $f(\vec{i},T)$
(dot line) as a function of $T$ in units $\tau_S$.}
\end{figure}

\section{Theoretical expectations}

\vskip 0.5cm

\indent

Let us first consider the phase shifts for the final interaction of the pions
in the Zel'dovich's approach \cite{zeld}, where the strong interactions
are evaluated, in the non relativistic limit, in terms of the s-wave phase
shifts for the three pairs of pions that can be formed. So, in the low
energy expansion, the relevant phase difference which will appear in our
calculations is
\be
\delta_S(1)-\delta_{MA}(2)=\frac{a_0}{7}\left[ 13 K_{+-} -\frac{K_{+0}+K_{-0}}
{2}+\frac{3}{2}(K_{+0}-K_{-0})\frac{\rho}{\Delta}\right],
\ee
where $a_0=0.2/m_\pi$ is the Weinberg's scattering length and we assume
$a_2=-2/7~a_0$ \cite{wein}; $K_{ij}$ is the momentum of the pion pairs in
their center of mass, reported in the Appendix A.
We list, for completeness, the expressions for all the phases:
\bea
\delta_S(1) &=& \frac{a_0}{7}\left[ 13 K_{+-}-2 (K_{+0}+K_{-0}) \right], \\
\delta_{MS}(1) &=& \frac{a_0}{7}\left[ 4 K_{+-}
-\frac{1}{2} (K_{+0}+K_{-0})- \frac{1}{2} (K_{+0}-K_{-0})
\frac{\Delta}{\rho}\right], \\
\delta_{MA}(2) &=& -\frac{3~a_0}{7}\left[
\frac{1}{2} (K_{+0}+K_{-0})+ \frac{1}{2} (K_{+0}-K_{-0})
\frac{\rho}{\Delta} \right].
\eea
The coefficients, involved in the expressions for $P$ and $S$, deduced
in the framework of the $\chi$PT at order $p^2$ and $p^4$ \cite{kambor}
are listed in table I.
As it is proved in \cite{kambor}, the values at order $p^2$ are
substantially changed by the corrections at the order $p^4$ in the energy
expansion.

We just considered the moduli given in ref.~\cite{kambor}
even though at $p^4$ in $\chi$PT all the coefficients acquire a small
imaginary part due to the loop contribution. This is justified as in
this paper we prefer to use the approach a' la Zel'dovich, in which all
the strong effects are included in the phase shifts.


\begin{table}
\centerline{TABLE I}
{\scriptsize
\begin{center}
\begin{tabular}{|c|c|c|c|c|c|c|} \hline
         &       &       &       &       &       &        \\
         &   a   &   b   &   c   &   d   &   e   &   f    \\
         &       &       &       &       &       &        \\ \hline
         &       &       &       &       &       &        \\
$O(p^2)$ & 0.699 & -4.55 & 0.540 &   -   &   -   &   -    \\
         &       &       &       &       &       &        \\ \hline
         &       &       &       &       &       &        \\
$O(p^4)$ & 0.842 & -7.30 & 0.750 & -3.78 & -9.46 & -0.721 \\
         &       &       &       &       &       &        \\ \hline
\end{tabular}
\end{center}}
\caption{Values of the coefficients of eq.~(20)
at order $p^2$ and $p^4$ in units $10^{-6}$.}
\end{table}


Now, we can predict the values of the coefficients for both $\Re(\ept)$
and $\Im(\ept)$ in eq.~(\ref{eq:nostra}), where only the leading terms in
the sin and cos expansion have been retained:
\be
\Gamma_\Delta (\kl) =8.4 \cdot 10^{-2}~\Re(\ept)
+1.6 \cdot 10^{-2}~\Im(\ept). \label{eq:nonum}
\ee
For $\Gamma [\kl \riar \f]$ we assume the experimental value in
ref.~\cite{pdg},
while for the coefficients $a$,..... we use the order $p^4$
values in table I, which would give rise to
\bea
Br(\kl \riar \f) &=& 12.4 \%, \label{eq:branch1} \\
Br(\kl \riar \pi^0 \pi^0 \pi^0) &=& 21.5 \%, \label{eq:branch2} \\
Br(\ks \riar \f) &=& 3.83 \cdot 10^{-5} \%, \label{eq:branch3}
\eea
to be compared with the experimental values
\bea
Br(\kl \riar \f) &=& (12.38 \pm 0.21) \%, \\
Br(\kl \riar \pi^0 \pi^0 \pi^0) &=& (21.6 \pm 0.8) \%, \\
Br(\ks \riar \f) &<& 4.9 \cdot 10^{-5} \%.
\eea
It is wise to stress that, while in all this paper we have not discarded
the strong phase shifts corrections and considered the exact Dalitz plot
contour, the values showed in the expressions (\ref{eq:branch1}),
(\ref{eq:branch2}), and (\ref{eq:branch3}) have been computed without
phases and in the non-relativistic approximation for the final pions,
in accordance with the lecterature. Indeed, adopting the first point of
view in computing the same branching ratios, one obtains
\bea
Br(\kl \riar \f) &=& 11.3 \%, \\
Br(\kl \riar \pi^0 \pi^0 \pi^0) &=& 19.5 \%, \\
Br(\ks \riar \f) &=& 3.20 \cdot 10^{-5} \%.
\eea
This result suggests a more careful analysis in fitting the coefficients
for $k \riar 3\pi$ decays, where one should take in account either the
dynamical effect of the strong phase shifts and the exact contour of the
Dalitz plot, which is plotted in fig.~2.

\begin{figure}
\vspace{5cm}
\caption{The contour of the available domain in the Dalitz plot for $\ko
(\kbo) \riar \f$, given by eq.~(61) (full line) and in the
non-relativistic approximation (dot line).}
\end{figure}

By using eq.~(\ref{eq:nonum}) and the values of $\Re (\ep)$ and $\Im (\ep)$
found in semileptonic decays and the moduli of the $\eta$ and
the value of {\it NA31} \cite{na31} for $\Re (\epp/\ep)$ we may
separate the contributions for $\Re (\ep)$, $\Im (\ep)$ and $|\epp|$ to
$\Gamma_\Delta (\kl)$:
\be
\Gamma_\Delta (\kl) =8.4 \cdot 10^{-2}~\Re(\ep)
+1.6 \cdot 10^{-2}~\Im(\ep)
+5.0 \cdot 10^{-1} |\epp|.
\ee
With the $10^{9} \kl$ expected at {\it DA$\Phi$NE} in one year, it will
certainly be possible measure a meaningful asymmetry coming from the real
part. In order to appreciate the contribution of the imaginary part, and
{\it a fortiori} of $\epp$ despite the enhancement factor
$\sqrt{2}\frac{\Re(A_0)}{\Re(A_2)}\sim 30$, a larger number of $\kl$ is
desired (larger luminosity or longer experiment).

In order to achieve a model independent
determination of $P$ and $S$ one needs to single out in the CP conserving
asymmetry $\Sigma^{+-0}_\Delta (T)$ the terms proportional to
$f(\vec{1},T)$ and $f(\vec{i},T)$. From the expressions given in
eqs.~(\ref{eq:pi}) and (\ref{eq:esse}) in the Appendix A and the values of
the integrals reported in the Appendix B one can realize that, with a few
percent approximation, $P$ and $S$ are given respectively by
\bea
P &=& a~c~F [\Delta^2 ], \\
S &=& a~c~F [\Delta^2 (\delta_S(1)-\delta_{MA}(2))],
\eea
and so it will be possible to extract the product $a\cdot c$ from $P$
and to test the final state interaction of the three pions from the
ratio $S/P$\footnote{We are grateful to Prof. H. Leutwyler for bringing
this point to our attention.}.

Since $a$ is experimentally well known from $\kl \riar 3 \pi$ decays
this seems a good, if not the best, way to get $c$, and consequently
$\Gamma (\ks \riar \f)$, which is difficult to be measured for the low
branching ratio and the necessity to separate it from the contamination of
$\kl$.

\section{Conclusions}

\vskip 0.5cm

\indent

We have been able to write the CP violating asymmetry in the energy of the
charged pions in the $\f$ decay of $\kl$ in terms of two parameters which
may be determined by studying the corresponding time-dependent asymmetry in
the decays of the neutral kaons with tagged initial strangeness.

With the values given for these parameters by $\chi PT$ and with the
evaluation of the final state interaction given by Zel'dovich one should
predict for the asymmetry the value
\be
\Gamma_\Delta (\kl) =8.4 \cdot 10^{-2}~\Re(\ep)
+1.6 \cdot 10^{-2}~\Im(\ep)
+5.0 \cdot 10^{-1} |\epp|. \label{eq:sogno}
\ee
within reach of the experiment {\it DA$\Phi$NE}, but with a small sensitivity
to $\Im (\ep)$ and even smaller to the value of $\epp$ indicated by {\it
NA31} and {\it E731} experiments.

The experimental study of the time dependence of the CP conserving
asymmetry $\Sigma^{+-0}_\Delta (T)$, needed to get the coefficients which
appear in eq.~(\ref{eq:sogno}), would supply the determination of $\Gamma
(\ks \riar \f)$, test the $\chi PT$ prediction and give information on
the final state interaction of the three pions.

\vskip 2cm
\centerline{\bf Acknowledgments}
\vskip 0.5cm
It's a great pleasure to thank Prof. H.
Leutwyler for clarifying conversations and valuable suggestions and
Dr. Gian Piero Mangano for useful discussions and continuous encouragement.

\vskip 2cm

\section*{Appendix A}

\vskip 0.5cm

\indent

The fundamental quantities defined by eq.~(\ref{eq:pes}), up to order $p^4$,
are
\bea
P &=& F \biggl[ a~c~\Delta^2 cos(\delta_S(1)-\delta_{MA}(2)) +
b~c~\Delta^2 \rho~cos(\delta_{MS}(1)-\delta_{MA}(2))+ \label{eq:pi} \\
&+& \left( (d+e)~c~\Delta^2 \rho^2 + (d-e)~c~\frac{\Delta^4}{3} \right)
cos~\delta_{MA}(2)+ a~f~\Delta^2 \rho~cos~\delta_{S}(1) + \nonumber \\
&+& b~f~\Delta^2 \rho^2 cos~\delta_{MS}(1)+ (d+e)~f~\Delta^2 \rho^3 +
(d-e)~f~\frac{\Delta^4}{3} \rho \biggr], \nonumber \\
S &=& F \biggl[ a~c~\Delta^2 sin(\delta_S(1)-\delta_{MA}(2)) +
b~c~\Delta^2 \rho~sin(\delta_{MS}(1)-\delta_{MA}(2))- \label{eq:esse} \\
&-& \left( (d+e)~c~\Delta^2 \rho^2 + (d-e)~c~\frac{\Delta^4}{3} \right)
sin~\delta_{MA}(2)+a~f~\Delta^2 \rho~sin~\delta_{S}(1) + \nonumber \\
&+& b~f~\Delta^2 \rho^2 sin~\delta_{MS}(1) \biggr]. \nonumber
\eea
The other terms do not contribute due to their odd parity in $\Delta$ and
$\rho$.
The phase space integrals involved in the previous expressions are shown in
the next Appendix.

Using eqs.~(\ref{eq:timeko}), (\ref{eq:timekbo}), (\ref{eq:gamt}), and
(\ref{eq:gambt}), as in ref.~\cite{sei}, we found
\bea
&\int ^{T}_{0}& \Gamma_{\Delta} (t)
(\overline{\Gamma}_{\Delta} (t))~~dt = {{1} \over
{2[1+{|\ept|}^{2} \pm 2\Re (\ept)  ]}}
\left\{ {{1-e^{-\Gamma_{S}T}} \over {\Gamma _{S}}}
 \biggl[
\Gamma_{\Delta} (\ku\riar\f) + \right. \\
&+& \left. |\ept|^{2}
\Gamma_{\Delta}(\kd\riar\f) + 2~~\Re\biggl( \ept
F[\Delta \cdot A^{*}(\ku\riar\f)~
A(\kd\riar\f)]\biggr)
\right] + \nonumber \\
&+& {{1- e^{-\Gamma_{L}T}} \over {\Gamma_{L}}} \cdot
\biggl[\Gamma_{\Delta}(\kd\riar\f) +
|\ept|^{2} \Gamma_{\Delta}(\ku\riar\f)+
\nonumber \\
&+& \left. 2~~\Re\biggl( \ept F[\Delta \cdot A^{*}(\kd\riar\f)~
A(\ku\riar\f)]\biggr)
\right]+ \nonumber
\\ &\pm& \Gamma_{\Delta}(\ku\riar\f)~
f( \vec{\ept},T) \pm
\Gamma_{\Delta}(\kd\riar\f)~
f( \vec{\ept^{*}},T) +
\nonumber \\ &\pm& \left[f\left(\overrightarrow
{F[\Delta \cdot A^{*}(\ku\riar\f)~
A(\kd\riar\f)]},T\right)
+ \right. \nonumber \\
&+& \left. \left. |\ept|^{2}f\left(\overrightarrow
{F[\Delta \cdot A(\ku\riar\f)~A^{*}
(\kd\riar\f)]},T\right) \right] \right\}, \nonumber
\eea
where the upper (lower) sign corresponds to $\ko$ ($\kbo$), while
once defined $\vec\mu \equiv (\Re (\mu), \Im (\mu))$, we have \cite{sei}
\be
f(\vec{\mu},T) \equiv{{2}\over{{\vec{\Gamma}}^{2}}}
[(1-\cos{\Delta mT}~{e}^{- {({\Gamma}_{S} + {\Gamma}_{L})T}\over{2} })
\vec{\mu} \cdot \vec{\Gamma}+
 (\vec{\mu} \wedge \vec{\Gamma})_{3}
\sin{\Delta mT}~{e}^{- {({\Gamma}_{S} + {\Gamma}_{L})T}\over{2}}],
\ee
with:
\bea
\Delta m &\equiv& {M_{L}-M_{S}}, \nonumber \\
\vec{\Gamma} &\equiv& {({{{\Gamma}_{S} + {\Gamma}_{L}}\over{2} },\Delta
m)}, \\
(\vec{\mu} \wedge \vec{\Gamma})_{3} &=& \Re (\mu) \Delta m - \Im (\mu)
{\Gamma_{S} + \Gamma_{L} \over 2}. \nonumber
\eea

The kinematical factors used in this paper are
\be
K_{ij} \equiv \left( {m_{\pi} \over 2} (2 Q - 3 T_k) \right)
^{{1 \over 2}},
\ee
where the $T_k$ are the kinetic energy of $\pi_k$ in the kaon rest frame
and
\bea
m_{\pi} &=& (m_{\pi^0}+2m_{\pi^+})/3, \\
Q &\equiv& m_{K}-m_{\pi^{0}}-2m_{\pi^{+}}=(83.562 \pm 0.032) MeV.
\eea

\section*{Appendix B}

\vskip 0.5cm

\indent

The Dalitz plot variables $r$ and $\phi$ are
\bea
T_{0} &\equiv& { Q \over 3 } ( 1 + rcos\phi ), \\
T_{\pm} &\equiv& { Q \over 3 } \left[ 1 + rcos \left( {2\pi \over 3} \mp
\phi \right) \right].
\eea

Given a function $h(r,\phi)$, we define
\be
F(h) \equiv
\frac{1}{(4\pi)^3 m_K}\frac{\sqrt{3}}{18} Q^2 \int \int r~dr~d\phi~
h(r,\phi)
\ee
(in particular $F(1)$ is the phase space factor).

The curve limiting the kinematically allowed region is, in the limit of
exact $SU(2)$ flavour symmetry,
\be
1 - (1 + \alpha ) r^2 - \alpha r^3 cos 3\phi = 0, \label{eq:impl}
\ee
with
\be
\alpha = { 2Qm_{K} \over {(2m_{K} - Q)^{2}}}.
\ee

The values of the integrals used in this paper are
\bea
F[1] &=& 1.954 \cdot 10^{-3} MeV, \nonumber \\
F[|\Delta|] &=& 7.694 \cdot 10^{-5} MeV, \nonumber \\
F[\rho^2] &=&  1.404 \cdot 10^{-6} MeV, \nonumber \\
F[\Delta^2] &=& 4.212 \cdot 10^{-6} MeV, \nonumber \\
F[\rho^3] &=&  -3.291 \cdot 10^{-9} MeV, \nonumber \\
F[\rho \Delta^2] &=& 9.873 \cdot 10^{-9} MeV, \nonumber \\
F[\rho^4] &=&  2.025 \cdot 10^{-9} MeV, \nonumber \\
F[\rho^2 \Delta^2 ] &=&  2.025 \cdot 10^{-9} MeV, \nonumber \\
F\left[\frac{\Delta^4}{3} \right] &=&  6.076 \cdot 10^{-9} MeV,
\nonumber \\
F\left[\frac{\rho\Delta^4}{3} \right] &=&  2.132 \cdot 10^{-11} MeV,
\nonumber \\
F[\rho^3\Delta^2] &=&  7.108 \cdot 10^{-12} MeV. \nonumber \\
F[\Delta^2 (\delta_S(1) - \delta_{MA}(2))] &=& 7.597 \cdot 10^{-7}
MeV, \nonumber \\
F[\rho \Delta^2 \delta_{S}(1)] &=&  -3.174 \cdot 10^{-9} MeV, \nonumber \\
F[\rho \Delta^2 (\delta_{MS}(1) - \delta_{MA}(2))]
&=& -8.995 \cdot 10^{-10} MeV, \nonumber \\
F[\rho^2 \Delta^2 \delta_{MS}(1)] &=& 8.390 \cdot 10^{-11} MeV, \nonumber \\
F[\rho^2 \Delta^2 \delta_{MA}(2)] &=&  -9.106 \cdot 10^{-11} MeV,
\nonumber \\
F\left[\frac{\Delta^4}{3} \delta_{MA}(2)\right] &=&
-2.732 \cdot 10^{-10} MeV, \nonumber
\eea
In order to compute them the contour equation~(\ref{eq:impl}) has been
numerically solved.

\pagebreak

\end{document}